\documentclass[%
 reprint,
 amsmath,amssymb,
 aps,
 pra,
]{revtex4-1}

\usepackage{graphicx}
\usepackage[space]{grffile}
\usepackage{latexsym}
\usepackage{textcomp}
\usepackage{longtable}
\usepackage{multirow,booktabs}
\usepackage{amsfonts,amsmath,amssymb}
\usepackage{natbib}
\usepackage{url}
\usepackage{hyperref}
\hypersetup{colorlinks=false,pdfborder={0 0 0}}
\newif\iflatexml\latexmlfalse
\usepackage[utf8]{inputenc}
\usepackage[main=english,german]{babel}
\usepackage{etoolbox}
\apptocmd{\sloppy}{\hbadness 10000\relax}{}{}

\newcommand\mathplus{+}

\begin{document}

\title{Optimal control with non-adiabatic Molecular Dynamics: application to the Coulomb explosion of Sodium clusters.}

\author{Adrián Gómez Pueyo}
\email{agomez@bifi.es}
\affiliation{Institute for Biocomputation and Physics of Complex Systems,
  University of Zaragoza, Calle Mariano Esquillor, 50018 Zaragoza, Spain}

\author{Jorge A. Budagosky M.}
\affiliation{Institute for Biocomputation and Physics of Complex Systems,
  University of Zaragoza, Calle Mariano Esquillor, 50018 Zaragoza, Spain}

\author{Alberto Castro}
\affiliation{Institute for Biocomputation and Physics of Complex Systems,
  University of Zaragoza, Calle Mariano Esquillor, 50018 Zaragoza, Spain}
\affiliation{ARAID Foundation, Calle Mar{\'{\i}}a Luna, 50018 Zaragoza, Spain}

\bibliographystyle{apsrev4-1}


\begin{abstract}
We present an implementation of optimal control theory for the
first-principles non-adiabatic Ehrenfest molecular dynamics model, which
describes a condensed matter system by considering classical point-particle
nuclei, and quantum electrons, handled in our case with time-dependent
density-functional theory. The scheme is demonstrated by optimizing the
Coulomb explosion of small sodium clusters: the algorithm is set to find the
optimal femtosecond laser pulses that disintegrate the clusters, for a given
total duration, fluence, and cut-off frequency. We describe the
numerical details and difficulties of the method.
\end{abstract}

\maketitle

\section{Introduction}

Thanks to the rapid development of laser technology~\cite{Keller_2003}, we
have nowadays access to high-intensity, short-duration pulse sources. This
availability has led to the birth of femtosecond science~\cite{Zewail_1994}
in the last decades of the 20th century, and of attosecond
science~\cite{Krausz_2009} in the new century. It has become possible to
study, in real time, the evolution of ions and electrons, as they evolve in
the influence of short and intense laser pulses. 

Theoretically, molecular dynamics (MD)~\cite{Rapaport_2004} studies the
processes that undergo condensed matter systems, in and out of equilibrium,
through simulations. For more than a few particles, it is impossible to
achieve a full exact quantum description of the problem. Often, the nuclei are
considered classical in an effort to alleviate the difficulties. In fact, the
name ``molecular dynamics'' is traditionally reserved for these models in which
the nuclear degrees of freedom are classical. Unfortunately, the remaining
electronic quantum problem in a non-equilibrium situation is still
very challenging~\cite{Stefanucci_2013}. Moreover, the electron-nuclear
coupling is in general non-adiabatic: in the adiabatic approximation, the
electronic system is fixed to the ground state corresponding to the
instantaneous nuclear configuration, which is obviously unsuitable for the
highly excited situation that high intensity sudden laser pulses lead to. For
a good description, one must use a first-principles theory for the
non-equilibrium many-electron quantum dynamics, non-adiabatically coupled to
the classic ionic movement through some form of MD.

One such scheme is the Ehrenfest molecular dynamics (EMD) in combination with
time-dependent density-functional theory
(TDDFT)~\cite{Runge_1984,Marques_2012}. 
EMD involves two approximations: first, one
separates the electronic and nuclear parts of the full wave function, arriving
at a ``time-dependent self-consistent model''~\cite{Gerber_1982}; second, the
short wave asymptotics idea~\cite{Wentzel_1926,Kramers_1926,Brillouin_1926} is used to take the
classical limit for the nuclear equations (see Ref.~\cite{Bornemann_1996} for
details on the justification of these steps). Then, one still needs to treat
the many-electron system, by making use of some electronic structure theory,
such as TDDFT. Ehrenfest MD based on TDDFT was first attempted by
Theilhaber~\cite{Theilhaber_1992} for external free-field problems, and its
applicability for laser-matter interaction was proved with various
examples~\cite{Saalmann_1996,Calvayrac_1998,Kunert_2001,Castro_2004,Miyamoto_2008,Russakoff_2015}.

The versatility of laser sources soon allowed the possibility of
\emph{controlling} quantum dynamics, a step beyond its mere use for
spectroscopy. The field of coherent or quantum
control~\cite{zewail1980,brumer1986,chan1991,chen1990,tannor1985,gaubatz1988,judson1992,assion1998,
  rice2000,shapiro2012,Brif_2010} was thus born.  A corresponding theoretical
framework had to be attached to the experimental advances. This is optimal
control theory (OCT), the mathematical branch that studies the ``inverse''
question in the study of dynamical systems: given a set of equations of motion
that depend on a set of parameters, OCT studies the methods to find the
parameters that optimize the system evolution. The first theoretical
calculations of OCT for quantum processes (QOCT) were reported in the
1980s~\cite{shi1988,Peirce_1988,Kosloff_1989,jakubetz1990}, and the field
quickly developed in the following decades~\cite{Werschnik_2007,Brif_2010}.

OCT has been demonstrated, for example, on the quantum equations for the
evolution of nuclear wave packets, with the aim of controlling nuclear
reactions. It was also combined with TDDFT~\cite{Castro_2012}, which opened
the road to the direct control of electronic systems. An obvious step forward
is the combination of OCT with an \emph{ab initio} non-adiabatic MD scheme, which
implies a control theory for a mixed quantum-classical system.  OCT for
classical or mixed quantum classical systems was already explored, for
example, in Refs~\cite{schwieters1993,botina1995}.  From a first-principles
perspective, the EMD-TDDFT scheme is a good candidate to attempt its
combination with OCT, due to its simplicity. In Ref.~\cite{Castro_2013} the
essential equations of an OCT for the EMD-TDDFT scheme were presented, while
Ref.~\cite{Castro_2016} demonstrated its numerical feasibility
for some simple two-electron molecules (H$_2$ and H$_3^+$). We have
further developed this numerical implementation, to allow for larger and more
complicated systems, and it is the purpose of this article to describe it in
detail, and show how it can be employed for some larger systems. In
particular, we have attempted the optimization of the Coulomb explosion of
small sodium clusters: the algorithm is set to find the optimal femtosecond
laser pulses that disintegrate the clusters, for a given total duration,
fluence, and cutoff frequency.

We have chosen the Coulomb explosion of sodium clusters because (1) it is a
far-from-equilibrium violent process~\cite{Fennel_2010} that requires
a nonperturbative scheme such as the one used here;
(2) the high intensity irradiation of simple metal clusters was
successfully studied with TDDFT in the
past~\cite{Calvayrac_2000,Wopperer_2015}; (3) the Coulomb explosion of these
systems was also already treated with
EMD-TDDFT~\cite{Calvayrac_1998,Suraud_2000,Ma_2001}, and it was shown how
there is an interesting interplay between the laser pulse, the electrons, and
the ionic motion.

Indeed, in principle the massive ionization that is responsible for the
Coulomb explosion may be achieved by tuning the laser frequency to the large
surface plasmon using the resonance-enhanced ionization mechanism. One could
then think that the explosion optimization problem can be simply solved by
making use of correctly tuned monochromatic pulses. However, the clusters
change during the pulse action in two different ways that complicate the
picture: (1) as ionization increases, the main resonance blueshifts, and (2) as
the ions start to move and separate from each other, the main resonance
red-shifts. It is therefore impossible to predict \emph{a priori} how to ``correct''
a simple, initially resonant, monochromatic pulse with other components, in
order to fully optimize the explosion process. This makes it a task suited for
OCT.

In Sec. \ref{QOCT} we present the theoretical framework which is at the
base of this work. In Sec. \ref{Results} we present the results obtained
for the Coulomb explosion of Na${}_{2}$, Na${}_{4}$ and Na${}_{8}$
clusters. In Sec. \ref{Conclusions} we discuss the implication of the
results found.  Finally, in the Appendix we detail the key aspects of the
computational implementation. Atomic units are used hereafter unless
stated otherwise.

\section{OCT for the EMD-TDDFT model}
\label{QOCT}

The details of the theory were given in Ref.~\cite{Castro_2013}; we only
outline here the key equations. An Ehrenfest system is a hybrid
quantum-classical system whose state is specified by a set of classical
conjugated position and momenta variables $\{R,P\}=\{\vec{R}_{\alpha}$,
$\vec{P}_{\alpha}\}^{K}_{\alpha=1}$ (where $\alpha$ runs over the $K$ nuclei
of our system, each with mass $M_{\alpha}$ and charge $z_{\alpha}$) and the
many-body wave function $\Psi$ of the $N$ electrons of our system. The forces
that determine the nuclear movement are given by:
\begin{align}
\nonumber
\vec{F}_{\alpha}[R(t),\Psi(t),u,t] = 
-\vec{\nabla}_{\alpha}W^{nn}(R(t))+z_{\alpha}\varepsilon(u,t)\vec{\pi}
\\
-\langle\Psi(t)|\nabla_{\alpha}\hat{H}[R(t),u,t]|\Psi(t)\rangle,
\label{eq:force}
\end{align}
where the nucleus-nucleus interaction $W^{nn}$ takes the usual Coulomb form
\begin{equation}
W^{nn}(R)=\sum_{\beta<\gamma}\frac{z_{\beta}z_{\gamma}}{|\vec{R}_{\beta}-\vec{R}_{\gamma}|}.
\end{equation}
The time-dependent function $\varepsilon(u,t)$ is the amplitude of the laser
pulse, which is polarized along the unit vector $\vec{\pi}$. The magnetic
component is ignored as we assume the dipole approximation and the length
gauge. $u=\{u_{1},u_{2},...,u_{M}\}$ is the set of parameters that can be
controlled and give the precise shape to the laser.
Finally, the last term in Eq.~\ref{eq:force} contains the electronic
Hamiltonian, given by
\begin{align}
\nonumber
\hat{H}[R,u,t] = 
\sum^{N}_{i=1}\frac{\hat{\vec{p}}^{2}_{i}}{2}+\sum_{i<j}\frac{1}{|\hat{\vec{r}}_{i}
- \hat{\vec{r}}_{j}|}
\\
+\varepsilon(u,t)\sum^{N}_{i=1}\hat{\vec{r}}_{i}\cdot\vec{\pi}
+ \sum^{N}_{i=1}\sum^{K}_{\beta}v^{\beta}(|\hat{\vec{r}}_{i}-\vec{R}_{\beta}(t)|),
\label{eq:hamiltonian}
\end{align}
where $\hat{\vec{r}}_{i}$ and $\hat{\vec{p}}_{i}$ are the position and
momentum operators for the $i$-th electron. The electron-nucleus
interaction in the last term is usually the Coulomb form
$v^{\beta}(r)=-\frac{z_{\beta}}{r}$,
but in practice we use pseudopotentials~\cite{Schwerdtfeger_2011}, which may
be nonlocal operators.

Rather than using the many-electron wave function, we want to use the TDDFT
Kohn-Sham (KS) formalism to substitute the real electronic system with an
equivalent one where the electrons do not interact with each other. This set
of electrons can be described using a single Slater determinant formed by $N$
spin orbitals. The time-dependent electronic densities of both systems are
identical by construction. It is given, in terms of the KS orbitals, by
\begin{equation}
n_t(\vec{r})\equiv n(\vec{r},t)=\sum^{N/2}_{m=1}2|\varphi_{m}(\vec{r},t)|^{2}.
\end{equation}
This assumes a spin-restricted situation in which we have an even number of
electrons, no magnetic fields, and no spin-orbit coupling. The system evolves
in a spin singlet and the spin orbitals are paired, each pair sharing the
same orbital part ($\varphi_m$) at all times. The examples presented below
assume this configuration.

The KS orbitals evolve according to the KS equations, a set of nonlinear Schr\selectlanguage{german}ö\selectlanguage{english}dinger-like equations:
\begin{equation}
i\frac{\rm d}{\rm dt}|\varphi_{i}(t)\rangle=\hat{H}_{\rm KS}[R(t),n_{t},u,t]|\varphi_{i}(t)\rangle\qquad(i=1,...,N/2).\label{KS}
\end{equation}
The KS Hamiltonian $\hat{H}_{\rm KS}$ is defined by
\begin{align}
\nonumber
\hat{H}_{\rm KS}[R(t),n_t,u,t] = 
\frac{1}{2}\hat{\vec{p}}^2 
+ \sum_{\beta} \hat{v}^\beta(\vert\hat{\vec{r}}-\vec{R}_\beta(t)\vert) 
\\
+ \varepsilon(u,t)\hat{\vec{r}}\cdot\vec{\pi}
+ \int\!\!{\rm d}^3r'\; \frac{n_t(\vec{r}')}{\vert \vec{r}-\vec{r}'\vert}
+ v_{\rm xc}[n_t](\vec{r})\,.
\label{eq:kshamiltonian}
\end{align}
The next-to-last term is the Hartree potential, and the last one is the
exchange-correlation potential. We assume an adiabatic approximation for this
last term, i.e., at each time $t$ it only depends on the density at that time
$n_{t}$. We have used, in particular, the adiabatic local density
approximation (ALDA), which is the first-step approximation to the intricate
problem of the exchange and correlation functional. However, the details of
the methodology presented here do not depend on the approximation used, and we
have preferred to use this generic choice because it is sufficient to draw
qualitative conclusions, and was successfully used in the past for
similar simulations~\cite{Suraud_2000}.

The force on each nucleus given in Eq.~(\ref{eq:force}) can be expressed as a
function of the time-dependent density, or alternatively, in terms of the KS
orbitals $\varphi\equiv\{\varphi_{m}\}^{N/2}_{m=1}$:
\begin{align}
\nonumber
\vec{F}_{\alpha}[R(t),\varphi(t),u,t] = 
-\vec{\nabla}_{\alpha}W^{nn}(R(t)) + 
z_{\alpha}\varepsilon(u,t)\vec{\pi}
\\
- \sum^{N/2}_{m=1}2\langle\varphi_{m}(t)|\vec{\nabla}_{\alpha}v^{\alpha}(|\hat{\vec{r}}
-\hat{\vec{R}}_{\alpha}(t)|)|\varphi_{m}(t)\rangle.
\label{eq:ksforce}
\end{align}
This fact permits us to avoid the full many-electron wave function completely and
is the basis of the EMD-TDDFT method. The full set of equations of motion
is given by Eqs.~(\ref{KS}) plus the usual classical Newton equations for the
nuclear movement:
\begin{equation}
\dot{\vec{R}}_\alpha(t) = \frac{1}{M_\alpha}\vec{P}_\alpha\,,
\label{eq:newton1}
\end{equation}
\begin{equation}
\dot{\vec{P}}_\alpha(t) = \vec{F}_\alpha[R(t),\varphi(t),u,t]\,.
\label{eq:newton2}
\end{equation}

Our purpose now is to look for the set of parameters $u$ that optimize the
behavior of the system with respect to a physical goal, e.g., the population
of some excited electronic state, the cleavage of a particular bond, or, as it
is the case of the examples shown below, the Coulomb explosion of a
cluster. This needs to be formulated as the maximization of a target
functional that depends on the system variables:
\begin{equation}
\mathcal{F}=\mathcal{F}[R(T),P(T),n_{T},u],
\end{equation}
where the time $T$ is the final time of the propagation interval $[0,T]$. This
definition depends only on the state of the system at the end of the
propagation interval (one could also formulate OCT for an evolution-dependent
target). 

Each set of parameters $u$ determines the evolution of the system,
$u\rightarrow R[u],P[u],\varphi[u]$,
so that the problem is reduced to the maximization of a function:
\begin{equation}
\label{eq:functiong}
G(u)=\mathcal{F}[R(T),P(T),n_{T},u].
\end{equation}
The algorithms for the maximization of functions are superior if they can make
use of the gradient of the function. One of the main results of OCT is the
derivation of an expression for the gradient; for the MD model that we are
discussing, the expression is~\cite{Castro_2013}
\begin{equation}
\label{eq:gradientfunction}
\frac{\partial G}{\partial u_{k}}=\int^{T}_{0}{\rm d}t\frac{\partial\varepsilon}{\partial u_{k}}(u,t)g(t)\,,
\end{equation}
\begin{equation}
g(t)=-\sum_{\beta}z_{\beta}\tilde{\vec{R}}_{\beta}(t)\cdot\vec{\pi}+2{\rm Im}\sum^{N/2}_{m=1}\langle\chi_{m}(t)|\hat{\vec{r}}\cdot\vec{\pi}|\varphi_{m}(t)\rangle\,.
\end{equation}
In this expression, there are new objects: new position variables $\tilde{\vec{R}}_{\beta}$ and one-particle orbitals $\{\chi_{m}\}^{N/2}_{m=1}$. Together with new momenta variables $\tilde{\vec{P}}_{\beta}$, they form the sometimes-called \emph{costate}, an auxiliary mixed quantum-classical system whose evolution is given by the following equations of motion:
\begin{align}
\dot{\tilde{\vec{R}}}_{\alpha}(t)=\frac{1}{M_{\alpha}}\tilde{\vec{P_{\alpha}}}(t),\label{CostR}
\\
\nonumber
\dot{\tilde{\vec{P}}}_{\alpha}(t) = 
\vec{\nabla}_{\alpha}\sum_{\beta}\tilde{\vec{R}}_{\beta}(t)\cdot\vec{F}_{\beta}[R(t),\varphi(t),u,t]
\\
+2{\rm Re}i\sum^{N/2}_{m=1}\langle\chi_{m}(t)|\vec{\nabla}_{\alpha}\hat{H}_{\rm KS}[R(t),n_{t},u,t]|\varphi_{m}\rangle,\label{CostP}
\\
\nonumber
|\dot{\chi}_{m}(t)\rangle = 
-i\hat{H}_{\rm KS}[R(t),u,t]|\varphi_{m}\rangle 
- i\sum^{N/2}_{n=1}
\hat{K}_{mn}[\varphi(t)]|\chi_{n}(t)\rangle
\\
-2\sum_{\beta}\tilde{\vec{R}}_{\beta}\cdot\vec{\nabla}_{\beta}\hat{v}^{\beta}
(|\hat{\vec{r}}-\vec{R}_{\beta}(t)|)|\varphi_{m}(t)\rangle\label{CostChi}\,.
\end{align}
The evolution equation for the costate orbitals $\chi_{m}$ contains a new, non-Hermitian term $\hat{K}_{mn}$, defined by
\begin{align}
\nonumber
\langle \vec{r} \vert \hat{K}_{mn}[\varphi(t)] \vert \chi_m(t)\rangle 
\\ = -4{\rm i}\varphi_m(\vec{r},t) {\rm Im} \int\!\!{\rm
  d}^3r'\chi_n(\vec{r},t)f_{\rm Hxc}[n_t](\vec{r},\vec{r}')\varphi_n(\vec{r},t)\,,
\end{align}
where the \emph{kernel} $f_{\rm Hxc}$ is the functional derivative of the Hartree and exchange and correlation potential functionals:
\begin{equation}
f_{\rm Hxc}[n](\vec{r},\vec{r}') = \frac{1}{\vert\vec{r}-\vec{r}'\vert} +
\frac{\delta v_{\rm xc}[n](\vec{r})}{\delta n(\vec{r}')}\,.
\end{equation}

Of course, the equations of motion for the costate need boundary conditions:
\begin{equation}
\tilde{\vec{R}}_{\alpha}(T)=-\frac{\partial}{\partial\vec{P}_{\alpha}}\mathcal{F}(R(T),P(T))\label{FCR},
\end{equation}
\begin{equation}
\tilde{\vec{P}}_{\alpha}(T)=\frac{\partial}{\partial\vec{R}_{\alpha}}\mathcal{F}(R(T),P(T))\label{FCP},
\end{equation}
\begin{equation}
\langle\vec{r}\vert\chi_{m}(T)\rangle=\frac{\delta\mathcal{F}}{\delta \varphi_m^*(\vec{r},T)}\,.\label{FCCHi}
\end{equation}
These are \emph{final} value conditions, and therefore the computation of the
$G$ gradient implies a forward propagation of the equations of
motion of our system and then a backwards propagation of the equations of
motion for the costate [Eqs.~(\ref{CostR}), (\ref{CostP}), and (\ref{CostChi})],
together with the previous final value conditions.

We have implemented these equations in the \texttt{OCTOPUS}
code~\cite{Marques_2003,Castro_2006}. The orbitals, densities, potentials,
etc. are represented in real space in a regular rectangular grid (for the
calculations shown below, the grid spacing was set to 0.8 a.u.) and the
system is contained in a simulation box $V$ (that was chosen to be spherical
for Na$_2$ and Na$_8$, and cylindrical for Na$_4$), with zero boundary
conditions in the edges. During the simulations, we add an imaginary absorbing
potential in the borders of the simulation box to simulate the ionization of
the electrons [see Eq.~(\ref{Vabs}) in the Appendix]. Essentially, the problem
consists of computing the maximum of the function $G$ defined above
[Eq.~(\ref{eq:functiong})], which we find with the help of the 
Broyden-Fletcher-Goldfarb-Shanno (BFGS)
algorithm~\cite{Nocedal_1980,Liu_1989}. The gradient necessitated by that
algorithm is obtained via the forward and backwards propagations mentioned
above, propagations that are done with the standard explicit fourth-order
Runge-Kutta scheme.

More details about the numerics are given in the Appendix, with particular
attention to the the key computational challenges posed by the integration of
OCT with the EMD-TDDFT method.


\section{Applications}
\label{Results}

We have applied the methodology described above to the following problem: find
the optimal laser pulses that Coulomb-explode the Na$_2$, Na$_4$, and Na$_8$
clusters for a given total duration, fluence, and cutoff
frequency. The first step is to fix an initial ground-state ionic
configuration. We have started from reference geometries found in the
literature~\cite{cccbdb,Solovyov_2002,Pal_2010} and have further relaxed them
with our own code in order to start the calculations with exactly
zero-force. As expected, the modified geometries are similar to the reference
ones, although we did find the known tendency of the local
density approximation (LDA) to underestimate bond lengths.

Next, we define a ``reference'' pulse in the following form: 
\begin{equation}
\label{eq:refpulse}
\varepsilon(t) = A e^{-(t-t_{0})^{2}/2\eta^{2}} \sin(\omega_0 t)\,.
\end{equation}
The frequency $\omega_0$ is set to 800~nm, the constant $t_{0}$ that
determines the position of the peak is set to half of the total pulse
duration, and the constant $\eta$ that determines the width is
set to 0.9$\tau_{0}$, where $\tau_{0}=2\pi/\omega_{0}$ is the period of the
laser pulse. The search space is then defined by fixing the total fluence to
that of this reference pulse, which can be tuned through the amplitude
constant $A$. The total duration $T$ of the pulses allowed in the search space
is set to 6, 12, or 24 periods $\tau_0$ of the reference frequency of 800~nm
($\approx 16, 32$, or 64~fs), depending on the case. We further impose a
cutoff frequency for the control function $\omega_{\rm cutoff}=0.5$ Ha, and
the search space is then parametrized as discussed in the Appendix, point
6. The optimization process is started with an initial random shape, and
in order to assure that the parameter space was adequately explored, we did
several runs for each case.

We then performed OCT runs in three different manners:
\begin{enumerate}

\item Using a target defined in terms of the classical momenta of the nuclei, i.e.,
\begin{equation}
\mathcal{F}[P(T)]=\sum_{i<j}|\vec{P}_{i}(T)-\vec{P}_{j}(T)|^{2},
\end{equation}
where $\vec{P}_{i}(T)$ is the momentum of the $i$-th nucleus at the end of the
propagation. The rationale behind this definition is clear: at the end of the
pulse, the ions in the cluster are required to have the maximum possible
opposing momenta. 

\item Using as a target an estimation for the ionization, i.e.,
\begin{equation}
\mathcal{F}[n_{T}] = -\int_V n_T(\vec{r})d^3r\,,
\end{equation}
where $n_T(\vec{r})$ is the electronic density at the end of the pulse action,
$t=T$. This amounts to the maximization of the electronic escape of the
system, which is what triggers the Coulomb explosion. First, we performed
optimizations with this target but fixing the nuclei to their equilibrium
position during the action of the pulse. This means that we are not, in this
second case, using the mixed quantum-classical OCT scheme stated in
Sec.~\ref{QOCT} but only the QOCT for TDDFT described in
Ref.~\cite{Castro_2012}.

\item Using the same target as in case 2, but letting the nuclei move freely
  during the optimization, i.e., using OCT on top of the full EMD-TDDFT model.

\end{enumerate}

Once we obtained the optimal pulses, we checked their performance by doing
time propagations with them for much longer times. We also used larger
simulation boxes for these test runs (which are the ones displayed below) to
make sure that the results were fully converged with respect to the box sizes
(given that ionization is the process that we are interested in) and to
allow the nuclei to travel far without reaching the simulation box
boundaries. Finally, we also performed equal test runs, making use of a
quasi-monochromatic laser pulse defined as the one in Eq.~(\ref{eq:refpulse}),
but tuned in each case to the most relevant excitation frequency of the
system. The goal was to check how the optimal pulses do a better job at
Coulomb-exploding the cluster than simple pulses tuned to make use of the
resonant-enhanced ionization phenomenon.

\subsection{Na${}_{2}$}

\begin{figure}
\begin{center}
\includegraphics[width=0.99\columnwidth]{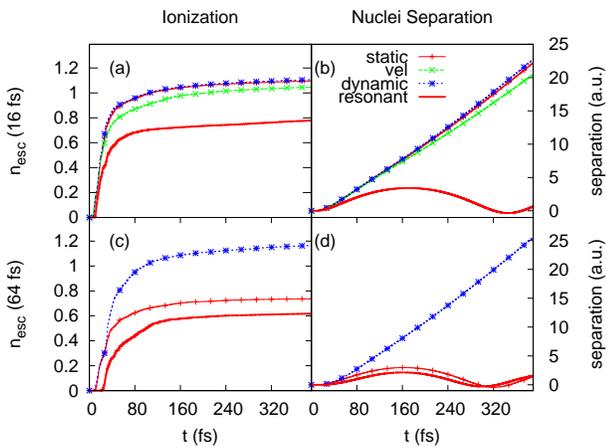}
\caption{{\label{na2-sep}
Left panels (a, c): Electrons escaped from the Na${}_{2}$ molecule after applying
    the pulses obtained using the three different optimization schemes: the
    momentum-based target (``vel''), the ionization target with fixed nuclei
    (``static''), and with moving nuclei (``dynamic''). We also plot the
    ionization obtained with the laser tuned to the resonance frequency pulse
    (``resonant''). Right panels (b, d): Separation between the nuclei from their
    equilibrium position. Top panels (a) and (b)
    correspond to 16~fs pulses; bottom panels (c) and (d) correspond to 64~fs pulses. %
}}

\end{center}
\end{figure}
\begin{figure}
\begin{center}
\includegraphics[width=0.99\columnwidth]{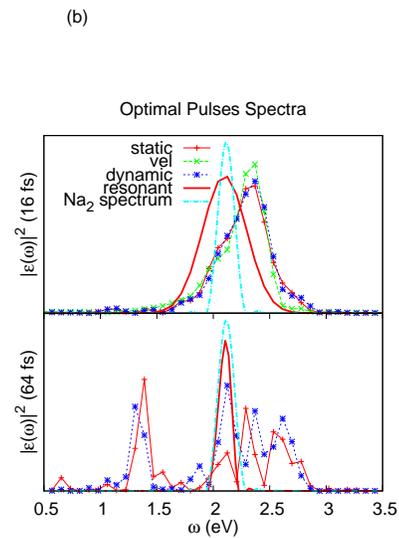}
\caption{{\label{na2-cpw}
Spectra of the optimized pulses and of the resonant pulse, and absorption
spectrum of the Na${}_{2}$ molecule for light polarized along its axis. Top
panel for the 16~fs pulses; bottom panel for the 64~fs pulses.%
}}
\end{center}
\end{figure}

The bond length of the Na$_2$  molecule is 5.82 a.u.~\cite{cccbdb}, but for
the simulations we have used the bond length computed by \texttt{OCTOPUS} within the
LDA approximation: 5.48 a.u., an underestimation which is to be expected for
the LDA. The polarization of the laser field was chosen to be parallel to the
dimer axis.


Fig.~\ref{na2-sep} shows the ionization induced in the molecule and the
separation between the nuclei obtained from the pulses found with each kind of
the three optimization schemes described above, and with a pulse in resonance
with the main excitation ($\omega=2.1$ eV) at the chosen polarization
direction. The top panels correspond to pulses with durations of $6\tau_0
\approx 16~$ fs, whereas the bottom panels are longer ones ($24\tau_0 \approx
64$~bs). For the former, we set the peak intensity of the reference pulse to
$10^{12}$~W cm$^{-2}$, and a lower intensity was used for the longer pulses
($3\times10^{11}$~W cm$^{-2}$).

As one can see, for the shorter pulses we have obtained Coulomb explosion for
all three optimization schemes. The frequency spectra of these pulses and the
absorption spectrum of the Na${}_{2}$ molecule (computed with the linear
response formalism of TDDFT, within the adiabatic LDA) are plotted in
Fig.~\ref{na2-cpw} (top panel for the short pulses). All the solutions found
are centered around the same frequency, slightly higher than the 2.1 eV of the
first resonance of the molecule. This blue-shift of the optimal frequency
should be blamed on ionization, that is significant already within the pulse
duration. In contrast, the nuclei do not move enough during the action of
these short 16~fs pulses to produce a significant effect. This can be learned
from the fact that the optimization performed with static nuclei provided a
similar solution, both in terms of the ionic movement and the shape of the
optimal pulses, than the two optimizations that do allow for ionic
movement. Finally, it can be seen how a laser tuned to the ground-state
resonance does not produce the photodissociation.

It is therefore clear that for those short pulses the ionic movement is not
very relevant, which is in line with previous research on the Coulomb
explosion of this kind of systems~\cite{Suraud_2000}. We therefore tried
longer pulses (24$\tau_0 \approx 64$~fs); the results are shown in the lower
panels of Figs~\ref{na2-sep} and \ref{na2-cpw}. We used the ``static'' and
``dynamic'' optimization targets, and it can be seen how while the former was
not capable of finding a dissociating pulse, the latter could. In other words,
the ionic movement was sufficiently relevant to make the clamped nuclei
approximation unsuitable, and therefore the full OCT+EMD+TDDFT combination was
necessary. This can also be seen in Fig.~\ref{na2-cpw}, which shows how the
optimal pulses obtained with the static and dynamic optimization schemes are
significantly different.

\subsection{Na${}_{4}$}

\begin{figure}
\begin{center}
\includegraphics[width=0.99\columnwidth]{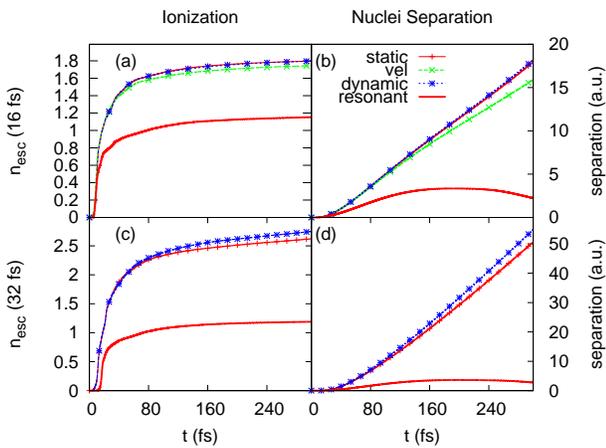}
\caption{{\label{na4-sep} Left panels (a, c): Electrons escaped from the
    Na${}_{4}$ cluster after applying the pulses obtained using the three
    different optimization schemes: the momentum-based target (``vel''), the
    ionization target with fixed nuclei (``static''), and with moving nuclei
    (``dynamic''). We also plot the ionization obtained with the laser tuned
    to the resonance frequency pulse (``resonant''). Right panels (b, d):
    Overall nuclear separation [Eq.~(\ref{overal-sep})].  Top panels (a) and (b)
    correspond to 16~fs pulses; bottom panels (c) and (d) correspond to 32~fs
    pulses. }}
\end{center}
\end{figure}

\begin{figure}
\begin{center}
\includegraphics[width=0.99\columnwidth]{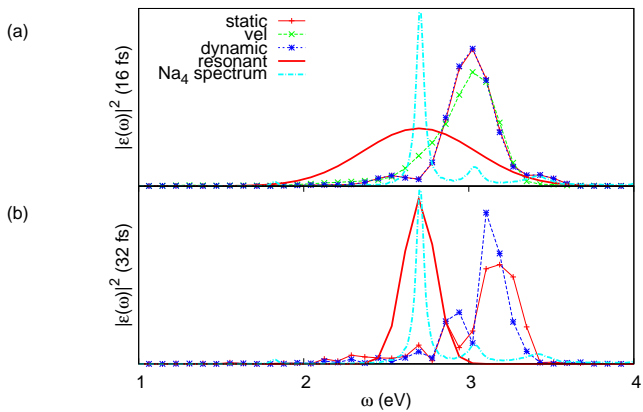}
\caption{{\label{na4-cpw} Spectra of the optimized pulses and of the resonant
    pulse, and absorption spectrum of the Na${}_{4}$ molecule for light
    polarized along the $X$ axis. Top panel for the 16~fs pulses; bottom panel
    for the 32~fs pulses.  }}
\end{center}
\end{figure}

The lowest energy Na${}_{4}$ isomer is planar, with the four atoms forming a
rhombus. We have attempted optimizations setting the polarization axis to
coincide with both the short axis ($X$) and the long axis ($Y$) of the rhombus. We
describe in detail only the former; for that direction, the main
resonance of the system is at $\omega_{X}=2.7$ eV. (For light polarized along
the $Y$ axis, the resonance is found at $\omega_{Y}=1.9$ eV.)


Fig.~\ref{na4-sep} shows the ionization and the overall nuclear separation,
defined as: 
\begin{equation}\label{overal-sep}
R(t)=\sum_{i<j}|\vec{R}_{i}(t)-\vec{R}_{j}(t)| - R_{\rm eq},
\end{equation}
where $R_{\rm eq}$ is the value of the previous sum at time zero (equilibrium
positions), so that $R(0)=0$.
The results plotted correspond to calculations with the laser
polarization along the $X$ axis. Fig~\ref{na4-cpw} shows the power spectrum of
the corresponding optimal pulses, along with the absorption spectrum of the
cluster for light polarized in the $X$ direction. 

We did calculations with 16~fs and 32~fs pulses, in order to assess the
relevance of the ionic motion. All optimization schemes produced optimal
shapes capable of Coulomb-exploding the cluster, both with the shorter and
with the longer pulses. For the shorter ones, the behavior of the system, and
the shape of the optimal pulse, are very similar with all optimization schemes,
including the static method in which the ionic movement is neglected. The
results start to be different with 32~fs pulses: although both the static and
dynamic schemes are successful, the shape of the optimal pulses (lower panel
of Fig.~\ref{na4-cpw}) are different, which shows that the ionic movement
starts to play a role. In all cases (short and long pulses, and with all
optimization schemes), the main peaks of the frequency distributions are once
again not over the resonance frequencies for the $X$ axis, but are slightly
blueshifted.


\subsection{Na${}_{8}$}

\begin{figure}
\begin{center}
\includegraphics[width=0.99\columnwidth]{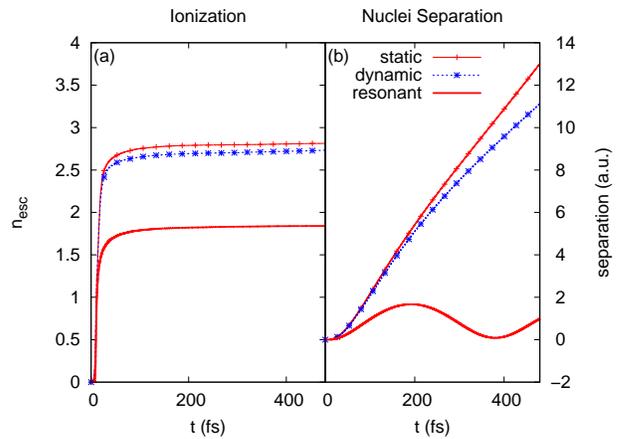}
\caption{{\label{na8-sep} (a) The ionization of the Na${}_{8}$ cluster after
    applying the pulses obtained using the static and dynamic nuclei
    ionization optimization schemes, using light polarized along the $Z$
    axis. (b) The overall separation between the nuclei [Eq.~(\ref{overal-sep})]
    that conform the Na${}_{8}$ cluster calculated with respect to the
    equilibrium separation.%
  }}
\end{center}
\end{figure}

\begin{figure}
\begin{center}
\includegraphics[width=0.99\columnwidth]{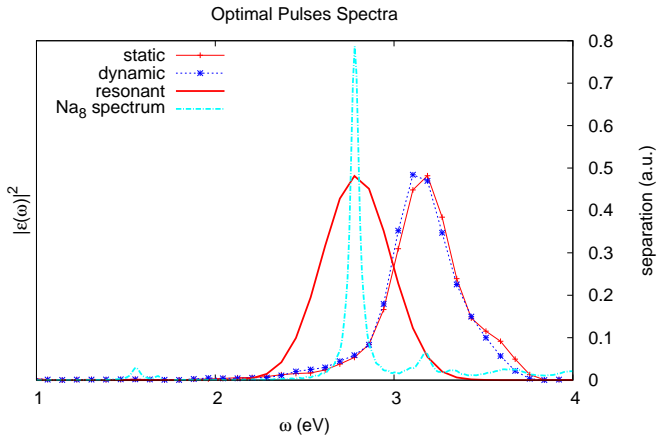}
\caption{{\label{na8-cpw} Spectra of the optimized pulses and of the resonant
    pulses, and absorption spectrum of the Na${}_{8}$ cluster for light
    polarized along the $Z$ axis.%
  }}
\end{center}
\end{figure}

Finally, we show results for the Na${}_{8}$ cluster. The equilibrium geometry
of this cluster has been the subject of some controversy~\cite{Pal_2010}, with
various candidates with D${}_{2d}$, D${}_{2h}$ or D${}_{4d}$ symmetries. We
have opted for the D${}_{4d}$ symmetry. This
configuration is associated with a strong absorption peak (a plasmon
resonance) at $2.55$ eV~\cite{Pal_2010}. We set the polarization direction of
the laser along the $Z$ axis. 


In this case, we have only used 16~fs pulses, due to the long computational
times needed for this bigger system. Fig.~\ref{na8-sep} shows the ionization
of the cluster for both the ``static'' and ``dynamic'' ionization optimization
schemes, and for the pulse at the resonance frequency (which in this case is
already a clear dominant plasmon peak). In the same figure, we show the
overall separation between the nuclei that conform the cluster, as defined in
Eq.~(\ref{overal-sep}). The use of both targets leads to successful
solutions. Interestingly, the ``static'' nuclei optimization slightly improves
the final ionization over the ``dynamic'' nuclei case. In both cases the
ionization curve saturates at around $~2.75$, whereas the ionization obtained
with the quasi-monochromatic resonance frequency pulse lies below two
electrons. In consequence, the nuclei oscillate only around their equilibrium
position for this last case.

The two optimal pulses obtained with the different schemes have similar
spectral composition, as can be seen in Fig.~\ref{na8-cpw}. Both are
blueshifted with respect to the monochromatic resonant pulse. Once again,
this shift must be attributed to the change in the plasmon resonance, as the
ionization takes place. Likewise, during the duration of the pulse, the ionic
movement is small, which is manifest from the fact that both static and
dynamic optimization schemes lead to similar results.

\section{Conclusions}
\label{Conclusions}

We have presented an implementation of OCT for the first-principles EMD-TDDFT
model, which describes a system by considering classical point-particle
nuclei, and quantum electrons handled with TDDFT. We have described the
details of its numerical implementation and have demonstrated its performance
by optimizing the Coulomb explosion of small sodium clusters: the algorithm
was set to find the optimal laser pulses capable of disintegrating the
clusters for a given total duration, fluence, and cut-off frequency.

In order to find those optimal pulses, we have used a standard gradient-based
nonlinear function maximization scheme for a merit function that values the
suitability of a given pulse. In order to compute the gradient, the scheme
needs consecutive forwards and backwards time propagations of the EMD-TDDFT
equations, and of some related, but more difficult equations of motion for a
fictitious auxiliary quantum-classical system. The main numerical difficulty
lies precisely within these propagations: (1) the need of accurate enough
gradients implies the need of a robust, yet expensive, propagator such as
fourth-order Runge-Kutta; (2) as we use absorbing boundaries in order to model
the electron escape from the cluster, in the backwards propagation the norm of
the KS orbitals increases and the propagation becomes unstable; (3) the
equations for the auxiliary fictitious system (the ``costate'') contain a term
that scales badly with the number of electrons (similarly to the exchange term
in time-dependent Hartree-Fock).

We have chosen some small Na clusters to show the functionality of the theory
and demonstrate how it can be used on larger and more complicated systems than
the hydrogen two-electron systems that were used before as a
proof of principle~\cite{Castro_2016}. The process optimized was the Coulomb
explosion of these clusters, a choice motivated by the interesting interplay
between laser pulse, electron ionization, and ionic movement that was found in
the past for these systems. It is to be expected that the explosion can be
helped through the resonance-enhanced ionization phenomenon, i.e., by tuning
the laser frequency to the plasmon resonance of the cluster. However, as the
laser ionizes the clusters, the main resonance of these cluster blueshifts,
and as the ions start to separate from each other, the same resonance
redshifts. It is not clear \emph{a priori} what frequency to use --or what
frequencies, for a pulse with complex structure.

It is therefore a problem suitable for an OCT calculation, as the optimal
solution cannot simply be guessed by intuition. We defined two optimization
targets: the opposing momenta between pairs of atoms at the end of the pulse,
and the number of escaped electrons -- the latter studied with both static and
dynamic ions. The scheme proved successful and we observed the following: (1) The
nuclei movement, for short 16~fs pulses, is not relevant, and the optimal
solutions can be found ignoring it, i.e., using OCT on top of electronic-only
TDDFT. (2) For longer 32~fs pulses, some differences between the optimal
pulses obtained with and without ionic movement start to appear, as shown in
the calculations for the Na$_4$ cluster. (3) For even longer 64~fs pulses, the
differences are substantial, and the OCT that ignores the ionic movement may
even not be capable of finding a successful shape, making the full OCT on top
of the nonadiabatic MD necessary. (4) The escape of electrons during the
action of the pulse, even for the shortest 16~fs ones, is sufficiently large
to shift the resonances, and therefore the optimal solutions do not
have their main peak at ground-state resonances (as one would expect from a
simple intuition based on the resonantly enhanced ionization idea) but at
slightly blueshifted values.

\appendix*

\section{Computational Aspects}
\label{CompAsp}

\begin{enumerate}

\item Optimization algorithms

  The problem of finding the maximum of a function depending on many variables
  is one of the most important in numerical analysis. Not surprisingly, there
  are a plethora of available methods, and their suitability will, of course,
  depend on factors such as the shape of the function (i.e., is it continuous? is it
  differentiable?), the number of parameters, etc. In the field of QOCT,
  several `\emph{ad hoc} algorithms were soon
  proposed~\cite{Zhu_1998,Zhu_1998a,Tannor_1992,Soml_i_1993,Maday_2003} and we
  have implemented a number of those in \texttt{OCTOPUS}. However, many of these lack
  generality: for example, they may assume a particular parametrization of
  the control function (in our case, the electric amplitude $\varepsilon(t)$),
  such as the full time-discretization of the function in the time axis (which
  can be understood as not parameterizing the control function at all). This
  fact makes it sometimes difficult to impose constraints on the shape of the
  functions, etc. Moreover, they are designed for pure \emph{quantum} OCT, and
  not suitable for the combined quantum-classical scheme discussed here.

Therefore, we have preferred to rely on general-purpose optimization
algorithms, such as
the low-storage BFGS algorithm~\cite{Nocedal_1980,Liu_1989}. 

\item Fourth-order accuracy in the gradient calculation

The function maximization will only proceed successfully with any algorithm if
the gradient is computed with sufficient accuracy. This is a challenging issue
for the present problem, since the gradient computation requires a complex and
long numerical procedure --the propagation of the differential equations shown
above, followed by the integration in Eq.~(\ref{eq:gradientfunction}). The key
numerical parameter is the time discretization step $\Delta t$. In our
experience, we found it necessary to perform all operations with $\Delta t^4$
order accuracy. Otherwise, the error escalates fast with increasing total
propagation time $T$. This implies time propagation algorithms of that order
(to be explained in the next point), and an integration of
Eq.~(\ref{eq:gradientfunction}) with a suitable order four scheme, such as
Simpson's rule, i.e. for $t_j = \Delta t j, (j = 0, 1, \dots, N)$:
\begin{equation}
\int_0^T\!\!{\rm d}t\; y(t) = \frac{\Delta t}{3}
\left[
y(0) + 2\sum_{j=1}^{N/2-1}y(t_{2j}) + 4\sum_{j=1}^{N/2}y(t_{2j-1}) + y(T)
\right]\,.
\end{equation}
With these precautions, the gradient computed with a time step $\Delta t$,
$\vec{\nabla}G_{\Delta t}$ will differ from the exact one $\vec{\nabla}G_0$ by
a fourth-order error in the time step:
\begin{equation}
\vert\vec{\nabla}G_{\Delta t} - \vec{\nabla}G_0\vert \sim O(\Delta t^4)\,.
\end{equation}

\item Propagation scheme.

  We have implemented various propagation schemes for the TDKS equations in
  \texttt{OCTOPUS}~\cite{Castro_2004a}. As mentioned above, a fourth-order integrator
  is required for these calculations. Note that one must not only propagate
  the KS system of electrons, but the full mixed quantum-classical system
  determined by the full set of variables for the real system state. We may
  denote by $Y=(R,P,\mathbf{\varphi)}$ to this full set of variables, where
  $R$ and $P$ are all the classical position and momenta variables, and
  $\mathbf{\varphi}$ are the KS orbitals.  For the costate,
  $Y=(\tilde{R},\tilde{P}, \mathbf{\chi})$. Generically speaking, we face the
  propagation of first-order nonlinear differential equations in the form:
\begin{equation}
\dot{Y} = f[Y(t),t]\,,
\end{equation}
where the dynamical function $f$ is given by the equations of motion above,
different for state and costate. A fourth-order accurate, all-purpose time
propagator is the classical explicit Runge-Kutta scheme (RK4):
\begin{eqnarray}
Y(t+\Delta t) & = & Y(t) + \frac{\Delta t}{6}( K_1 + 2K_2+2K_3+K_4)\,,
\\
K_1 & = & f(Y(t),t)\,,
\\
K_2 & = & f(Y(t)+\Delta t/2K_1,t+\Delta t/2)\,,
\\
K_3 & = & f(Y(t)+\Delta t/2K_2,t+\Delta t/2)\,,
\\
K_4 & = & f(Y(t)+\Delta tK_3,t+\Delta t)\,.
\end{eqnarray}
Each time step, therefore, requires four evaluations of $f$. For the state,
this implies the generation of the KS Hamiltonian for the given nuclear
configuration and electronic density, its application on the set of KS
orbitals, and the computation of the force, given in Eq.~(\ref{eq:ksforce}).
The costate propagation is more complicated. First of all, note that the
equation for the costate orbitals $\chi_m$ [Eq.~(\ref{CostChi})] contains two
extra terms, in addition to the KS Hamiltonian: the second term in the right-hand
side consists of the application of the operators
$\hat{K}_{mn}[\varphi(t)]$. These are nonlocal operators, similar to the
non-local operators used in Hartree-Fock theory. One needs to apply one of
these operators for each Kohn-Sham orbital, and this fact makes the co-state
propagation to scale badly with the system size (roughly fourth order with
the number of electrons), in contrast to the KS orbital propagation, which in
principle is only second order. Moreover, the $\hat{K}_{mn}[\varphi(t)]$
operators are not Hermitian, due to the presence of the imaginary part in
their definition. Finally, the last term in the right-hand side makes the
equation inhomogeneous.

The ``real'' state $(R,P,\varphi)$ could be propagated by making use of any of
various methods suitable to propagate Kohn-Sham equations (such as the ones
described in Ref.~\cite{Castro_2004a}), in combination with, for example, the
standard velocity-Verlet algorithm for the classical ions. However, due to the
extra difficulties mentioned above, we have found that for the costate
$(\tilde{R},\tilde{P},\chi)$ the robust all-purpose explicit RK4 scheme is the
best option, despite its extra computational cost.

\item Non-Hermitian evolution.

The equation of motion for the $\chi_m$ orbitals contains some non-Hermitian
operators. This implies that the equations do not preserve the orbital norms
--also because of the presence of one inhomogeneous term. For the purpose of
the optimization shown below, moreover, we add a non-Hermitian term to the KS
Hamiltonian of Eq.~(\ref{eq:kshamiltonian}): an absorbing boundary potential
that is used to account, in an approximate manner, for the possibility of
ionization. The idea is to split the simulation region into an inner region,
in which the absorbing potential is zero and an outer region in which the
action of the potential, due to its imaginary value, removes electron
charge. In our calculations, the absorbing region has width $L$ and the
absorbing potential definition is given by
\begin{equation}
V_{\rm abs}(\vec{r}) = i\eta \sin^2\left(\frac{d(\vec{r})\pi}{2L}\right)\,,\label{Vabs}
\end{equation}
where $d(\vec{r})$ is the distance from point to $\vec{r}$ to the frontier
between inner and outer simulation box regions. The integral of the electron
density in the simulation region,
\begin{equation}
N(t) = \int_V\!\!{\rm d}^3r\;n(\vec{r},t)\,,
\end{equation}
is no longer a constant of motion, and the electron loss may serve to estimate
the ionization probability of the process.

Note that the propagation
of the co-state requires the prior knowledge of the true state evolution. One
could perform the forward propagation of the state and store it at all
times. This storage should be on disk because of its enormous size. The
input and output operations required are too slow, and therefore the best solution
is actually to recompute the system state at all times by propagating it
backwards along with the costate. 



When marching forwards, the norm of the KS orbitals decreases due to the
presence of the absorbing boundaries, and this makes the propagation a stable
numerical procedure (since the error is proportional to the norm). However,
when marching backwards, the norm increases, and the evolution is
unstable. This can only be cured by establishing some milestone points during
the forward propagation at which the orbitals are stored: the orbitals propagated
backwards are compared to the orbitals stored at those points, and if they
differ by an unacceptable amount, the program stops with an error; otherwise,
the backwards-propagated orbitals are substituted by the stored ones.

\item Operator derivatives

The equations of motion contain various terms of the form:
\begin{equation}
\vec{\nabla}_{{\bf R}_a} f(\hat{\vec{r}}-\vec{R}_a)\,,
\end{equation}
i.e. derivatives of operators that depend on the ionic positions. Numerically,
in our real-space approach used by the \texttt{OCTOPUS} code, it is advantageous to use
for their computation the identity:
\begin{equation}
\vec{\nabla}_{\vec{R}_a} f(\hat{\vec{r}}-\vec{R}_a) = 
-i\left[
\hat{\vec{p}},f(\hat{\vec{r}}-\vec{R}_a)
\right]\,.
\end{equation}
This permits us to substitute the numerical derivatives of the function $f$ by
numerical derivatives of the wave functions (see Ref.~\cite{Hirose_2005} for a
discussion on these issues). However, Eq.~(\ref{CostP}) also contains the
derivatives of the forces
\begin{equation}
\vec{\nabla}_{\alpha}\sum_{\beta}\tilde{\vec{R}}_{\beta}(t)\cdot\vec{F}_{\beta}[R(t),\varphi(t),u,t]\,,
\end{equation}
which are, in fact, double derivatives with respect to the potential
function. For these terms, we have employed a finite-difference scheme,
i.e., the derivatives are estimated by computing the forces at neighboring
values of the nuclear positions.

\item Control function parametrizations

We allow for various possible parametrizations $\varepsilon =
\varepsilon(u_1,\dots,u_M,t)$ in our optimal control implementation in the
\texttt{OCTOPUS} code. One simple possibility is using directly the real-time
discretization
\begin{equation}
u_j = \varepsilon(t_j)\,.
\end{equation}
However, this implies a large number of parameters. Moreover, it is hard to
establish constraints on the function, such as a cut-off frequency, for
example. For the examples shown below, we have used the parametrization
described in Ref.~\cite{Krieger_2011}, based on a Fourier expansion:
\begin{equation}
\varepsilon(u,t) = 
\sum_{n=0}^{K} \left[ a_n(u) \cos\left(\frac{2\pi}{T}n t\right) + b_n(u) \sin\left(\frac{2\pi}{T}n t\right) \right] \,.
\end{equation}
This form naturally sets a cutoff frequency on the shape of the control
function, $\omega_{\rm max} = \frac{2\pi}{T}K$. The parameters $u$ are a set
of hyperspherical angles that run over all possible Fourier coefficients
$a_n,b_n$, subject to the following constraints:
\begin{equation}
\varepsilon(u,0) = \varepsilon(u,T) = 0\,,
\end{equation}
\begin{equation}
\int_0^T\!\!{\rm d}t\;\varepsilon(u,t) = 0\,,
\end{equation}
\begin{equation}
\int_0^T\!\!{\rm d}t\;\varepsilon^2(u,t) = F_0\,.
\end{equation}
The first two conditions are natural restrictions for an electric field
produced by a laser pulse. The last condition establishes a ``fixed-fluence''
condition: the search is done over all possible laser pulses whose intensity
integrated in time (the fluence) is given. A condition over the energy of the
pulse such as this one is necessary for our optimization target, since
otherwise the obvious solution to Coulomb-explode a cluster would be a pulse
with infinite intensity.

\end{enumerate}

\begin{acknowledgments}
We acknowledge support from the Ministerio de Econom{\'{\i}}a y Competitividad
(MINECO) through Grants No. FIS2013-46159-C3-2-P and No. FIS2014-61301-EXP.
\end{acknowledgments}

\bibliography{biblio}


\end{document}